\documentclass{amsproc}
\usepackage{amsmath, amssymb, amsfonts, amstext, verbatim, amsthm}

\input xy
\xyoption{all}
\usepackage{setspace}
\usepackage{enumerate}
\usepackage{prettyref}
\newrefformat{prop}{Proposition~\ref{#1}}
\newrefformat{cor}{Corollary~\ref{#1}}
\newrefformat{thm}{Theorem~\ref{#1}}
\newrefformat{lem}{Lemma~\ref{#1}}
\newrefformat{def}{Definition~\ref{#1}}

\theoremstyle{plain}
\newtheorem{thm}{Theorem}[section]
\newtheorem{lem}[thm]{Lemma}

\newtheorem{quest}[thm]{Question}

\theoremstyle{definition}
\newtheorem{rem}[thm]{Remark}
\newtheorem{defn}[thm]{Definition}
\newtheorem{ex}[thm]{Example}

\newcommand{\op}[1]{\!\!\mathop{\rm ~#1}\nolimits}

\begin{document}

\title{Curved String topology and Tangential Fukaya Categories}
\author{Daniel Pomerleano}
\maketitle

\section{Introduction}
    In this paper, we construct new examples of two-dimensional TQFTs over the prop $C_*(\overline{\mathcal{M}_{g,n}})$. Our primary methods are algebraic: we make use of the well known theorem of Kontsevich and Soibelman \cite{KS} that \vskip 10 pt \begin{em} Given a compact and smooth $\mathbb{Z}/2\mathbb{Z}$ graded Calabi-Yau $A_{\infty}$ algebra $B$ for which the Hodge to De-Rham spectral sequence degenerates, a choice of splitting for this spectral sequence gives rise to a TQFT \end{em} \vskip 10 pt 

   For a compact, smooth, Calabi-Yau variety, a (dg-version of) the derived category of quasicoherent sheaves $QCoh(\mathcal{X})$ satisfes all of the above conditions. Homological Mirror Symmetry \cite{Kontsevich} predicts that the associated TQFT is expected to be equivalent to Gromov-Witten TQFT on the mirror CY variety $\mathcal{X}^{\vee}$. Now consider $\mathcal{Y}$ to be a smooth but non-compact Calabi-Yau variety. Then $QCoh(\mathcal{Y})$ is a non-compact Calabi-Yau category, and by a modified version of the theorem of Kontsevich and Soibelman, we can get a so-called \begin{em} positive-output \end{em} TQFT. The Landau-Ginzburg model uses deformation theory to compactify these theories by deforming the above category by a superpotential $w$, which is an algebraic function with a proper critical set. Recent work \cite{Preygel,Lin-Pomerleano} shows that this gives rise to a TQFT. \vskip 10 pt 
   
  Similarly, there is a positive output TQFT called \begin{em} string topology \end{em} for a compact oriented manifold $\mathcal{Q}$ associated to the dg-category of dg-modules  $D(C_*(\Omega \mathcal{Q}))$ over the dg algebra $C_*(\Omega \mathcal{Q})$ \cite{Lurie}, where $\Omega \mathcal{Q}$ denotes the based loop space of $\mathcal{Q}$ at some arbitrary point. Throughout this paper, all coefficients are taken to be $\mathbb{C}$, the field of complex numbers. As we explain below, this category is a smooth but not compact category. The relationship with string topology is revealed by the following calculation for the Hochschild homology:
  
 $$ \mathbf{HH}_*(C_*(\Omega \mathcal{Q}))\cong C_*(\mathcal{LQ})$$

  There is a natural compact CY category associated to such a manifold, the category of perfect modules over $C^*(\mathcal{Q})$, which however is not smooth. Such categories give rise to TQFT's with \begin{em} positive-input\end{em}. When $\mathcal{Q}$ is simply connected, these two algebras are related via Koszul duality. Namely, the following isomorphisms hold: \vskip 10 pt

$$\mathbb{R}Hom_{C^*(\mathcal{Q})}(\mathbb{C},\mathbb{C})\cong C_*(\Omega \mathcal{Q})$$  $$C^*(\mathcal{Q})\cong \mathbb{R}Hom_{C_*(\Omega \mathcal{Q})}(\mathbb{C},\mathbb{C})$$

 and in fact this gives rise to fully faithful functors:
 $$perf(C_*(\Omega\mathcal{Q}) \to D(C^*(\mathcal{Q}))$$ and $$perf(C^*(\mathcal{Q})^{op}) \to D(C_*(\Omega \mathcal{Q})^{op})$$

Here $perf(C_*(\Omega\mathcal{Q})$ or $perf(C^*(\mathcal{Q})^{op})$ denotes the subcategory of perfect modules, which is defined for the reader below. Nevertheless, $\mathbb{C}$ is not a compact generator in the category $D(C_*(\Omega \mathcal{Q}))$ which means that Koszul duality does not give rise to an equivalence of the full derived categories. The starting point for this work is that if $\mathcal{Q}$ is $T^n= S^1\times S^1\times \cdots \times S^1$, Dyckerhoff \cite{Dyckerhoff} proved the following theorem: 

\begin{thm} Let w be a function on $\mathbb{C}[[x_1,x_2,\ldots,x_n]]$ with isolated singularities. The object $\mathbb{C}$ is a compact generator for $MF(\mathbb{C}[[x_1,x_2,\ldots,x_n]],w)$. Otherwise stated, $Hom_{MF(\mathbb{C}[[x_1,x_2,\ldots,x_n]],w)}(\mathbb{C},-)$ defines an equivalence of categories: $$ MF(\mathbb{C}[[x_1,x_2,\ldots,x_n]],w) \to D(Hom_{MF(\mathbb{C}[[x_1,x_2,\ldots,x_n]],w)}(\mathbb{C},\mathbb{C})) $$ \end{thm} 

Here $MF$ denotes the category of matrix factorizations, whose definition occupies much of section 2. The relationship between this theorem and the previous discussion is that $C_*(\Omega T^n)$ is isomorphic to $\mathbb{C}[z_1,z_1^{-1},z_2,z_2^{-1},\ldots z_n,z_n^{-1}]$, the Laurent polynomial ring in several variables. As $T^n= S^1\times S^1\ldots \times S^1$ is not simply connected, we complete at the augmentation ideal of this ring to obtain $\mathbb{C}[[x_1,x_2,\ldots,x_n]]$. In such cases, $MF(\mathbb{C}[[x_1,x_2,\ldots,x_n]],w)$ defines a quantum field theory. This result can be viewed as a deformed Koszul duality in the sense that $Hom_{MF(\mathbb{C}[[x_1,x_2,\ldots,x_n]],w)}(\mathbb{C},\mathbb{C}) \cong H^*(T^n)$ with a deformed $A_{\infty}$ structure $m$.  \vskip 10 pt

 In this paper, we will consider simply connected manifolds $\mathcal{Q}$ whose minimal models are pure Sullivan algebras (again we will review this terminology). The first part of our paper makes precise and then gives an answer to the following question:
 
\begin{quest} If $C^*(\mathcal{Q})$ is a pure Sullivan algebra and given an element $w\in Z(C_*(\Omega \mathcal{Q}))$, when is $\mathbb{C}$ a compact generator of $MF(C_*(\Omega \mathcal{Q}),w)$ defining an equivalence with $D(H^*(C^*(\mathcal{Q})),m)$? \end{quest} 
 
 We will examine our condition in the special case that the differential of our pure Sullivan algebra is quadratic. As mentioned earlier, morally, one can think of a potential $w$ as ``compactifying" the field theory. In the final section, inspired by a program of \cite{Seidel}, we explain how the simplest of our theories, such as when $\mathcal{Q}=\mathbb{C}P^n$ can be interpreted as geometric compactifications of the cotangent bundle $T^*CP^n$ inside of a certain root stack. For the latest update in the relationship between Fukaya categories of $T^*CP^n$ and string topology, the reader should consult \cite{Abouzaid}. \vskip 10 pt 
 
 This paper is a summary of a short talk given in June 2011 at the String-Math conference. A forthcoming paper \cite{Pomerleano} will develop further the ideas discussed herein while developing some singularity theory that is suggested by analogy with the commutative case. The author would like to thank his advisor Constantin Teleman for suggesting the possibility of transporting ideas from the Landau-Ginzburg model to String topology as well as for his support and guidance throughout this project. The author would also like to thank Mohammmed Abouzaid, Denis Auroux for help with the symplectic geometry and Toly Preygel for teaching me about curved algebras. I have learned a lot of what I know so far about this subject from them. 
 
 \section{Background and Algebraic Setup}
Recall that a dg-module (or $A_{\infty}$-module) $N$ over a dg-algebra $A$ (or $A_{\infty}$-algebra)  is \begin{em}perfect\end{em} if it is contained in the smallest idempotent-closed triangulated subcategory of $Ho(A)$ generated by $A$.

\begin{defn} A dg-algebra $A$ over $\mathbb{C}$ is \begin{em} compact \end{em} if $A$ is perfect as a $\mathbb{C}$ module (in this special case this simply says that $A$ is equivalent to a finite dimensional vector space). A dg-algebra $A$ is \begin{em}smooth\end{em} if $A$ is perfect as an $A-A$ bimodule.\end{defn} 

A very useful criterion for smoothness is given by the notion of finite-type of Toen and Vaquie \cite{Toen2}. 
 
 \begin{defn} A dg-algebra $A$ is of finite type if it is a homotopy retract in the homotopy category of dg-algebras of a free algebra $(\mathbb{C}\langle v_1,v_2,\ldots,v_n\rangle,d)$ with  $dv_j \in \mathbb{C}\langle v_1,v_2,\ldots,v_{j-1}\rangle$ \end{defn}
 
 \begin{lem} If $A$ is of finite type then $A$ is smooth. The converse is also true if $A$ is assumed to be compact. \end{lem}
 
 \begin{lem} With the notation of the previous section, the dg-algebra $C_*(\Omega \mathcal{Q})$ is smooth.\end{lem}
In the simply connected case, this follows from the classical Adams-Hilton construction \cite{Adams} and the above theorem of To\"en-Vaquie. \vskip 10 pt 

\section{Pure Sullivan algebras and Curved algebras} 

We consider Pure Sullivan dg-algebras $\mathcal{B}$ of the form: 
 $$ (\wedge V, d)=(\mathbb{C}[x_1,...x_n]\otimes \bigwedge(\beta_1,...\beta_m), d(\beta_i)=f_i(x_1,\ldots, x_n),d(x_j)=0) $$ 
where the $deg(x_i)$ are even and negative, the functions $f_i$ have no linear term, and the $deg(\beta_i)$ are odd $>1$. 
We further assume that $dim(H^*(\mathcal{B}))<\infty$. \vskip 10 pt

From a field-theoretic point of view, it is important to note that $\mathcal{B}$ is in particular elliptic and hence $H^*(\mathcal{B})$ is a Poincare duality algebra \cite{Felix}. Because the deformation theory of $C_{\infty}$ algebras and Frobenius $C_{\infty}$ algebras is known to coincide, $\mathcal{B}$ has a natural Calabi-Yau structure. The general theory of Koszul duality in turn implies that its Koszul dual $\mathcal{A}$ has a non-compact Calabi-Yau structure.\vskip 10 pt

The above cochain algebras determines canonically an $L_{\infty}$ model, $\mathfrak{g}$ = $\pi_*(\Omega(\mathcal{M}))\otimes \mathbb{C}$ for the space $\mathcal{Q}$. Using the homological perturbation lemma, we have an explicit $A_{\infty}$ model for $\mathcal{A}=C_*(\Omega \mathcal{Q})$ of the form 

$$(Sym(\mathfrak{g}_{even}) \otimes \Lambda(\mathfrak{g}_{odd}),m)$$

 A formula for the higher multiplications appears in \cite{Baranovsky}, but the key facts are as follows. First, the strict morphism of the abelian Lie algebra $\pi_{even}(\Omega(Q)) \to \mathfrak{g}$ corresponds to the inclusion of $Sym(\mathfrak{g}_{even})\cong \mathbb{C}[u_1,\ldots,u_m] \to \mathcal{A}$. The higher multiplications $m_n$ are multi-linear in these variables for $n \geq 3$. Finally, the $A_{\infty}$ algebra is strictly unital and the augmentation 
 $U\mathfrak{g} \to \mathbb{C}$ is also a strict morphism. \vskip 10 pt
 
 The reader should be warned that in the presence of quadratic terms in the $f_i$, the above identification with $Sym(\mathfrak{g}_{even}) \otimes \Lambda(\mathfrak{g}_{odd})$ is only an identification of vector spaces. In other words, there can be a non-trivial Lie bracket $B:\mathfrak{g}_{odd}\otimes \mathfrak{g}_{odd} \to \mathfrak{g}_{even}$, which means that forgetting higher products, $U\mathfrak{g}$ is a Clifford algebra over $Sym(\mathfrak{g}_{even})$.  It also seems worth pointing out that the even variables $u_i$ can be thought of as being Koszul dual to the odd variables $\beta_i$. Meanwhile the variables in $\mathfrak{g}_{odd}$, from here on denoted as $e_j$, are dual to the even variables $x_j$ above.  \vskip 10 pt
 
  For example, if $\mathcal{Q}=\mathbb{C}P^n$, we have the following specific model: 
 $$U(\mathfrak{g})= \mathbb{C}[u] \otimes \Lambda(e), m_{n+1}(e,e,e,\ldots,e)=u$$ We can then consider potentials of the form $w=u^d$. \vskip 10 pt

Next, we discuss how to define an appropriate category of matrix factorizations. This section adopts the ideas of the foundational work \cite{Preygel} to our non-commutative context. For concreteness, let us consider as before the above $A_{\infty}$-algebra $\mathcal{A}$, and an element $w \in \mathbb{C}[u_1,\ldots, u_m]$ of degree $2j-2$. We define a variable $x$ of degree $2j-2$. The element $w$ defines a mapping from $$w:\mathbb{C}[x] \to \mathcal{A}$$, and we can consider the $A_{\infty}$ algebra $\mathcal{A}_0=(\mathcal{A}[e], de=w)$, where $e$ now has degree $2j-1$. \vskip 10 pt

\begin{defn} We define $Pre(MF(\mathcal{A},w))$, to be the full subcategory of $mod(\mathcal{A}_0)$ consisting of modules which are perfect over $\mathcal{A}$. \end{defn}
  This category has a natural $\mathbb{C}[[t]]$(degree $t=-2n$) linear structure because it is acted on by the $\mathbb{C}$-finite modules $D_{fin}(\mathbb{C}[e]/e^2)$. By Koszul duality, this latter category is equivalent to the category $perf(C[[t]])$. 
 \vskip 10 pt 
 
There is also a deformation theoretic interpretation of the above action. We note that the element $tw$ also defines a Maurer-Cartan element in $\mathbf{HH}^*(\mathcal{A},\mathcal{A})[[t]]$. Such a Maurer-Cartan solution allows us to twist the differential on $$(\bigoplus_n \mathcal{A}^{\otimes n}[[t]],d_{\mathcal{A}}) $$ by the differential determined by the formula:

\[td_w: a_0 \otimes a_1 \otimes \cdots \otimes a_n \mapsto \sum_{i=0}^{n-1} (-1)^{i+1} t a_0 \otimes a_1 \otimes \cdots \otimes a_i \otimes W \otimes a_{i+1} \otimes \cdots \otimes a_n. \]

giving rise to a topological coalgebra 

$$\mathbf{C}=(\bigoplus \mathcal{A}^{\otimes n}[[t]],d_{\mathcal{A}}+td_w)$$.
 
 \begin{lem} The functor $ M \to ((\bigoplus B^i(\mathcal{A})\otimes M)[[t]],d_{M/A}+te\wedge)$ defines a fully faithful functor:
 $$Pre(MF(\mathcal{A},w)) \to D(\mathbf{C}-comod)$$ \end{lem} 

Finally, we define $$MF(\mathcal{A},w) = Pre(MF(\mathcal{A},w))\otimes_{\mathbb{C}[[t]]}\mathbb{C}((t)) \cong Pre(MF(\mathcal{A},w))/Perf(\mathcal{A}_0) \cong D(\mathbf{C}-comod)\otimes_{\mathbb{C}[[t]]}\mathbb{C}((t)) $$ 

It is often convenient to work with the formal Ind-completion $Ind(MF(\mathcal{A},w))$ which we shall denote by $MF^{\infty}(\mathcal{A},w)$.

We have constructed a category of curved modules for a curved $A_{\infty}$ algebra which arises as a deformation of an uncurved $A_{\infty}$ algebra. It is worth pointing out that there is a more general notion of a curved $A_{\infty}$ algebra, a notion which is most developed in the case of dg-algebras. \vskip 10 pt

\begin{defn} A triple $ B=(A,w,d)$ consisting of a $\mathbb{Z}/2\mathbb{Z}$ graded algebra $A$, a function of even degree $w$, and an derivation $d$ of odd degree is called a graded curved dg-algebra if $d^2=[w,a]$ \end{defn}\vskip 10 pt

 \begin{defn} A (left) curved module over a curved dg-algebra is a $\mathbb{Z}/2\mathbb{Z}$ graded (left) module over $A$ together with an odd derivation $d$ such that $d^2=w.$ \end{defn}
 
  There is a $\mathbb{Z}/2\mathbb{Z}$ graded dg-category of modules, which we denote by $B-mod$. Positselski has studied curved Koszul duality extensively and in particular defined various versions of the derived category of curved modules over a curved dg-algebra. In particular, he considers:
 
  \begin{defn} We denote by $B-proj$ the $\mathbb{Z}/2\mathbb{Z}$ graded dg-subcategory of $B-mod$ consisting of modules $M$ whose underlying graded modules $M^{\sharp}$ are projective. We define $Ho(B-proj)$ to be its homotopy category. \end{defn} 


In many cases of interest, $B-proj$ coincides with the categories defined previously and is sometimes convenient to work with. In the case that $A$ is a commutative ring and $w$ is a non-zero function, $B-proj$ is nothing but the usual category of matrix factorizations.\vskip 10 pt

For the case of a general curved $A_{\infty}$ algebra, it is a bit unclear how to construct an interesting triangulated category of modules. One possible definition is to consider topological modules over the completed bar coalgebra $\prod B^i(A)$. This amounts to considering only those modules such that $l_n: A^{\otimes n}\otimes M \to M$ vanish in sufficiently high degree and further imposing the analogous condition for morphisms between two modules \cite{Positselski}. 

\section{ The criterion for properness and coformal $\mathcal{Q}$}

In this section we discuss a criterion for smoothness and properness of the category $MF(\mathcal{A},w)$. To state the criterion, we must consider the category of curved bimodules $$MF(\mathcal{A}\otimes \mathcal{A}^{op},w\otimes1-1\otimes w)$$ and we define $\mathbf{HH}^*(MF(\mathcal{A},w))$ to be $Hom_{MF(\mathcal{A}\otimes \mathcal{A}^{op},w\otimes1-1\otimes w)}(\mathcal{A},\mathcal{A})$. Using either description of our category, this can be computed explicitly as: 

$$\mathbf{HH}^*(MF(\mathcal{A},w))  \cong (\mathbf{HH}^*(\mathcal{A},\mathcal{A})((t)),d_{\mathcal{A}}+[tw,]) $$

The following is the analogue of Dyckerhoff's theorem for our situation:

\begin{thm} If $(\mathbf{HH}^*(\mathcal{A},\mathcal{A})((t)),d_A+[tw,])$ is finite over $\mathbb{C}((t))$, then $\mathbb{C}$ generates the category $MF(\mathcal{A},w)$.\end{thm} 

(Sketch of proof) We have an action of $\mathbf{D}=\mathbb{C}[u_1,\ldots,u_m]$ on $MF(\mathcal{A},w)$ which factors through the complex 
$\mathbf{HH}^*(MF(\mathcal{A},w))$. For any u in $\mathbf{D}$, we let $K_u(\mathbf{D})$ be the diagram
\[
\xymatrix {\mathbf{D}\ar[r]^u & \mathbf{D}}
\]

For the sequence $\bar{u}=(u_1,\ldots,u_m)$ we define $$ K_{\bar{u}}(\mathbf{D}) = \otimes K_{u_i}(\mathbf{D})$$ and we consider the colimit of the obvious diagram:

$$ K_{\bar{u}}(\mathbf{D}) \to K_{\bar{u}^2}(\mathbf{D})\to K_{\bar{u}^3}(\mathbf{D}) \ldots $$

which we denote by E. For any object $\mathbf{O}$ in $MF(\mathcal{A},w)$,  we have an augmentation $$E\otimes \mathbf{O} \to \mathbf{O} \to cone(e)$$

Because the action of $\mathbf{D}$ factors as above, we can conclude that $cone(e)$ is zero and that this map is an isomorphism. Now the objects $K_{\bar{u}^i}(\mathbf{D})\otimes \mathbf{O}$ are in the triangulated subcategory generated by $\mathbb{C}$ because their cohomologies are finite. Because $\mathbf{O}$ is compact and can be expressed as a colimit of $K_{\bar{u}^i}(\mathbf{D})\otimes \mathbf{O}$, we have that $\mathbf{O}$ is a direct summand of one of the $K_{\bar{u}^i}(\mathbf{D})\otimes \mathbf{O}$ generated by $\mathbb{C}$ as well. \vskip 10 pt

We denote by $RHom_c(MF^{\infty}(\mathcal{A},w),MF^{\infty}(\mathcal{A},w))$ the category of continuous endofunctors in the sense of \cite{Toen}. Similarly to the works \cite{Preygel,Lin-Pomerleano}, we can apply our generation result to the category to prove the following fact which implies smoothness for  $MF(\mathcal{A},w)$:

\begin{thm} $RHom_c(MF^{\infty}(\mathcal{A},w),MF^{\infty}(\mathcal{A},w)) \cong MF^{\infty}(\mathcal{A}\otimes \mathcal{A}^{op}, w\otimes1- 1\otimes w)$ \end{thm}

We can make this condition more tractable by considering the deformation theory of the pure Sullivan algebra $A^{!}$ itself. As noted in the introduction, for any simply connected space of finite type, we have fully faithful functors induced by the $C^*(\mathcal{Q})-C_*(\Omega \mathcal{Q})$ bimodule $\mathbb{C}$. It then follows from a result of Keller \cite{Keller} that for such a fully faithful functor there is a canonical equivalence in the homotopy category of $B(\infty)$ algebras:
 
 $$\mathbf{HH}^*(C^*(\mathcal{Q}),C^*(\mathcal{Q}))\cong \mathbf{HH}^*(C_*(\Omega \mathcal{Q}),C_*(\Omega \mathcal{Q}))$$
 
In particular these two Koszul dual algebras have equivalent deformation theories. Suppose a commutative dga has a free-commutative model $(\bigwedge V,d)$ where $V$ is a finite dimensional vector space. There is a very explicit complex quasi-isomorphic as a dg-Lie algebra to $\mathbf{HH}^*((\bigwedge V,d),(\bigwedge V,d))$. Recall that $T^{poly}(V)$ is the Lie-algebra of polyvector fields on $\bigwedge V$ with Schouten bracket. Part of Kontsevich's formality theorem says that the HKR map: $$T^{poly}(V) \to \mathbf{HH}^*(\bigwedge V)$$ is the first Taylor coefficient in an $L_{\infty}$ quasi-isomorphism between the two.\vskip 10 pt

We can think of the derivation $d$ as corresponding to a vector-field $v$. It follows from a spectral sequence argument that the HKR map gives a quasi-isomorphism:

$$(T^{poly}(V),[v,-]) \to \mathbf{HH}^*((\bigwedge V,d),(\bigwedge V,d))$$

\begin{lem} This map can be corrected to an $L_{\infty}$ quasi-isomorphism. In the case of a pure Sullivan algebra, the first Taylor coefficient agrees with the HKR map.\end{lem}
 
In the pure Sullivan case, potentials tw in $\mathbf{HH}^*(\mathcal{A},\mathcal{A})[[t]]$ correspond to odd-polyvector fields 
$tw(d/de_1,d/de_2, \ldots d/de_m) \in T^{poly}(\mathcal{B})[[t]]$. After passing to the generic fiber, the Hochschild cohomology is given by :
$$(T^{poly}(V)((t)),[v+tw(d/de_1,\ldots,d/de_m),])$$

\begin{defn} By analogy with the case of ordinary matrix factorizations, we will say that $w$ has an isolated singularity if the homology of this complex is finite dimensional. \end{defn}

Let $\mathcal{B}$ be a pure Sullivan algebra, whose Lie model $\mathfrak{g}$ is formal. We have that $U\mathfrak{g}$ is a graded Clifford algebra over 
$\mathbb{C}[u_1,\ldots,u_m]$. We let $D_k$ be the closed subvariety of $\mathbb{C}[u_1,\ldots u_m]$ for which $ rank(B)\leq{k}$ and assume further that the $D_k-D_{k-1}$ is smooth. Let $R$ denote $U\mathfrak{g}/(w)$. In this setting we can relate our notion of isolated singularity to another possible notion of isolated singularity in non-commutative geometry.

\begin{thm} Let $\mathcal{B}$ be a pure Sullivan algebra, whose Lie model $\mathfrak{g}$ is formal and as above. Let $w$ be a potential which intersects the varieties $D_k$ transversally at every point. Then: \begin{enumerate}[(a)] \item $w$ has isolated singularities 
                                                     
                                                       \item $Proj(R)$ has finite homological dimension as an abelian category.
                                                       \end{enumerate}  \end{thm}

The first statement is a calculation, so we explain the second one. Consider the exact functor between derived categories 

$$\pi: D^b(Gr-R) \to D^b(Proj(R)) $$

We can consider the abelian subcategory of $Gr-R$, denoted $Gr-R_{\geq i}$ which consists of modules $M$ such that $M_p=0$ for $p \leq i$ Restricted to this subcategory, 

$$\pi_{\geq i}: D^b_{\geq i}(Gr-R) \to D^b(Proj(R)) $$ 

has a right adjoint 

$$R\omega:D^b(Proj(R)) \to D^b_{\geq i}(Gr-R)$$

Thus we will show that for any $ M,N \in D^b(Gr-R)$, $\op{Ext}^i(M,R\omega \circ \pi(N))$ vanishes for large $i$. \vskip 10 pt

Suppose that $Q$ is a graded prime ideal different from the maximal ideal and lying in a component of $D_k$, but not $D_{k-1}$. Now denote by $P$ the prime ideal corresponding to the irreducible component of $D_k$ which $Q$ is in. One can prove that the correspondence $P \mapsto rad(PR)$ gives a bijection between (graded) prime ideals in $\mathbb{C}[u_1,\ldots,u_m]$ and (graded) prime ideals of $U\mathfrak{g}$. We have a short exact sequence $$0\to S \to R/(rad(PR),Q) \to R/rad(QR) \to 0$$ where $S$ is $R/rad(QR)$ torsion by the assumption that the prime $Q$ lie in a component of $D_k$ but not $D_{k-1}$. Now we know by our condition, that $\mathbb{C}[u_1,\ldots,u_m]/Q[l]$ has a finite resolution as a $\mathbb{C}[u_1,\ldots,u_m]/P$ module and thus so does $R/(rad(PR),Q)$[l] as a $R/rad(PR)$ module. 

\vskip 10 pt The above exact sequence reveals that $\op{Ext}^i_{R/(rad(PR))}(R/rad(QR)[l],M)$ is $R/rad(QR)$ torsion for sufficiently large $i$. It is also easy to show from the transversality hypothesis that $R/rad(PR)[l]$ has finite homological dimension over $R$. Now the following lemma \cite{Brown} and the change of ring spectral sequence enable us to conclude the result:

\begin{lem} Let $R$ be a graded FBN ring. Given a bounded complex C in $D(Gr-R)$ if $\op{Ext}^i(R/P[l],C)$ is $R/P$ torsion for $i>>j$ for every two-sided prime ideal $P$ then $\op{Ext}^i(M,C)$ vanishes for $i>>0$.\end{lem}

The proof is as in the stated reference provided that we note the Gabriel correspondence between minimal injectives and graded prime ideals for graded FBN rings and that every bounded complex in $Gr-R$ is equivalent to a minimal complex of injectives. \vskip 10 pt

\begin{ex} For $\prod S^{2n_j}$ the condition that $w$ has an isolated singularity is similar to the usual Jacobian condition and states that $\mathbb{C}[u_1,\ldots, u_m]/(u_idw/du_i)$ be finite dimensional. \end{ex}

The field theory assigned to $S^2$  can be computed explicitly. Working with $\mathbb{Z}/2\mathbb{Z}$ gradings for ease of notation, we have the following calculations:

\begin{lem}The endomorphisms algebra $End(\mathbb{C},\mathbb{C})$, in the category of curved modules over $(\mathbb{C}[x],x^{2d})$, deg(x)=1, is given by $\mathbb{C}[e]/e^2$, with one higher multiplication $m_{2d}(e,e...e)=1$ \end{lem} This calculation is very similar to \cite{Dyckerhoff} theorem 4.7. The deformed algebra has an obvious cyclically symmetric inner product given by Poincare duality. \vskip 10 pt
 
The calculations below are tedious but straightforward for the patient reader. 

\begin{lem} The Hochschild homology in this case is given by even elements $e_0,e_1,...e_{2d-1}$. The pairing on $\mathbf{HH}_*$ given by the above TQFT is given by $\langle e_i,e_j \rangle$ is non-zero if i+j=2d-1 and zero otherwise. \end{lem}
\begin{lem} The Hochschild cohomology of this TQFT is concentrated in even degree and equals $$\mathbb{C}[e]/e^2=1, \text{if d=1}$$ $$\mathbb{C}[e,v]/(e^2=0, ev=0, v^{d}=0),    \text{otherwise} $$ \end{lem} 

\section{Tangential Fukaya categories}
Given the close connection between TQFTs and Floer theory, in this section we aim to give a Floer theoretic interpretation of the previous sections in some special cases. For motivation, let us consider the easiest case of a symplectic mirror to a Landau-Ginzburg model, that of $S^2$. We think of a sphere as being the (open) disk bundle of the cotangent bundle, $D^*(S^1)$, compactified by the points at 0 and $\infty$. This is then mirror to $(C_*(\Omega(S^1)) \cong \mathbb{C}[z,z^{-1}], w= z+1/z)$.\vskip 10 pt

If we wish to understand the mirror to the Landau-Ginzburg model $(\mathbb{C}[z,z^{-1}], w= z^d+1/z^d)$ we can either consider the Fukaya category of the orbifold $S^2//(\mathbb{Z}/d\mathbb{Z})$, where $\mathbb{Z}/d\mathbb{Z}$ acts by rotations that fix the two points, or more concretely a Fukaya category where we require disks to intersect the compactifying divisor with ramification of order $d$. If one wants to generalize this to higher dimensional projective spaces, the mirror of $\mathbb{C}P^n$ is well known to be the Landau-Ginzburg model: $$(z_0+ z_1 +\ldots z_n ,\text{ } z_0z_1\ldots z_n =1)$$ Mirror symmetry predicts that to obtain the mirror manifold to: $$(z_0^{d_0}+ z_1^{d_1} +\ldots z_n^{d_n}, z_0z_1\ldots z_n =1) $$  one performs the root stack construction, to be defined below, along the toric divisors $z_i=0$. \vskip 10 pt 

Given a variety $\mathcal{X}$ and a collection of effective Cartier divisor $\mathcal{D}_i$, and $d_i$ a collection of positive integers. The Cartier divisors define a natural morphism $$\mathcal{X}\to [\mathbb{A}^n/(\mathbb{C}^*)^n]$$ The root stack $\mathcal{X}_{(\mathcal{D}_i,d_i)}$ is defined to be the fibre product $$ \mathcal{X}\times_{[\mathbb{A}^n/(\mathbb{C}^*)^n]}[\mathbb{A}^n/(\mathbb{C}^*)^n]$$ where the map $[\mathbb{A}^n/(\mathbb{C}^*)^n]\to [\mathbb{A}^n/(\mathbb{C}^*)^n]$ is the $d_i$-power map. The root stack defines an orbifold, which has non-trivial orbifold stabilizers along the divisors (the coarse moduli space is exactly $\mathcal{X}$ and away from $\mathcal{D}_i$ the map $\mathcal{X}_{(\mathcal{D}_i,d_i)} \to \mathcal{X}$ is an isomorphism). The main property is that to give a map into the root stack is equivalent to giving a map into $\mathcal{X}$ which is ramified to order $d_i$ along the divisors $\mathcal{D}_i$. The formulas in the previous paragraph easily generalize to produce mirrors to toric Fano manifolds with root constructions performed along toric divisors. \vskip 10 pt

The is also a clear symplectic interpretation of the TQFT associated in section 3 to $\mathcal{Q}=\mathbb{C}P^n$ when $w=u$. Namely, we have an anti-holomorphic involution of $I:\mathbb{C}P^n\times \mathbb{C}P^n \to \mathbb{C}P^n \times \mathbb{C}P^n$ given by $(z,w) \to (\bar{w},\bar{z})$. Its fixed point set: $ \mathcal{L} : \mathbb{C}P^n \to \mathbb{C}P^n \times \mathbb{C}P^n$, is a Lagrangian submanifold. We have that $HF^*(\mathcal{L},\mathcal{L}) \cong \mathbb{C}((t))[e]/(e^{n+1}=t)$ and in this case, the category $MF(\mathcal{A},w)$ is isomorphic to the full subcategory of the Fukaya category of $\mathbb{C}P^n\times \mathbb{C}P^n$ split-generated by $\mathcal{L}$. \vskip 10 pt

In what follows it will be important to think of $\mathbb{C}P^n\times \mathbb{C}P^n$ as a symplectic cut. $T^*\mathcal{Q} - \mathcal{Q}$ acquires a Hamiltonian $S^1$ action by rotating the geodesics (which then give rise to Reeb orbits when restricted to the unit cotangent bundle). This induces a natural Hamiltonian action on $(T^*\mathcal{Q}-\mathcal{Q}) \times \mathbb{C}$. The moment map got this Hamiltonian $S^1$ action 

 $$(T^*\mathcal{Q}-\mathcal{Q}) \times \mathbb{C} \to \mathbb{R}$$

is given by 
 
$$(x,z) \mapsto H(x)+1/2|z|^2$$ 

Where $H(x)=|x|$ is the Hamiltonian associated to the Hamiltonian action on $T^*\mathcal{Q} - \mathcal{Q}$. We then take the reduced space, that is the preimage of a regular value quotiented out by the $S^1$ action. Finally, we glue back in the zero section to obtain a manifold $\mathcal{X}$ which is a symplectic compactification of the open disk bundle $D^*(\mathcal{Q})$ by the smooth divisor $\mathcal{D}$. Concretely, the divisor is defined by the equation 

$$ \sum z_iw_i=0$$ 

 The Lagrangian $\mathcal{L}$ corresponds to the zero section. The above discussion suggests that our other deformations should be realized by performing the $d$-th root stack construction on $\mathbb{C}P^n \times \mathbb{C}P^n$ along $\mathcal{D}$ or in this case equivalently counting holomorphic disks with a prescribed tangency to the divisor $\mathcal{D}$. It is worth noting that the above construction can be applied to other manifolds with periodic geodesics e.g., $S^n$, $n>1$, or $\mathbb{H}P^n$. In the case of $S^n$, one obtains $S^n$ as a Lagrangian submanifold of the projective quadric $Q_n$. The symplectic compactifying divisor is isomorphic to $Q_{n-1}$. One can repeat the calculations below and a similar picture develops to that described in this section. \vskip 10 pt

 We now define the Floer homology that we wish to consider. The tangency Floer theory we define here will agree with the Floer theory in the root stack because curves with components which live entirely in the divisor $\mathcal{D}$ occurs in real codimension $\geq 2$.  
  
  We want to consider the moduli space of holomorphic disks  $$ f: (D^2,S^1) \to (\mathbb{C}P^n\times \mathbb{C}P^n,\mathcal{L})$$  
  $$ \text{if  } f(p)\in \mathcal{D}, \text{then  } m(p)=d $$ 
 
  Here $m(p)$ denotes the intersection multiplicity, which we require to be exactly $d$ at each point of intersection. We consider the \cite{FOOO} compactification of this moduli space and compactify our moduli space as a subspace of $\mathfrak{M}_{k}(\mathbb{C}P^n \times \mathbb{C}P^n,\mathcal{L})$ in the obvious way. We will denote the moduli space by $\mathfrak{M}_{j,d,k}$, where $j$ denotes the number of intersection points with $\mathcal{D}$ and $d$ the multiplicity. As in \cite{FOOO}, we consider some model for chains on $C_*(\mathbb{C}P^n,\mathbb{C}((t)))$ and using the evaluation maps $$ev_i: \mathfrak{M}_{j,d,k+1} \to C_*(\mathbb{C}P^n)$$ to define a sequence of higher products $$ m_k(\alpha_1,...\alpha_k)=\sum ev_{0,*}(\prod ev_i^*(\alpha_i))t^{n}$$
 
 One can show that the standard complex structure $J$ can be perturbed in the complement of a neighborhood of $\mathcal{D}$ to be regular for this moduli problem. We wish to prove:
 
 \begin{thm} $ HF^*(\mathcal{L},\mathcal{L}) \cong End_{MF(\mathcal{A},w)}(\mathbb{C},\mathbb{C})$, where $(\mathcal{A},w)$ is the curved algebra associated in section three to $\mathcal{Q}=\mathbb{C}P^n$, with potential $w=u^d$. \end{thm}
 
  The key lemma is:  
 
\begin{lem} The Hochschild class the infinitesimal deformation class defined on $H^*(\mathbb{C}P^n,\mathbb{C}[t]/t^2)$ using the moduli space $\mathfrak{M}_{1,d,k}$ is gauge equivalent to the $d$-fold cup product of the Hochschild class determined by $\mathfrak{M}_{1,1,k}$ \end{lem}
To prove this result, we proceed by induction and consider a certain codimension one submanifold of the moduli space of disks with two points of intersection with the divisor, one of multiplicity $d-1$ and one simple intersection. Analysis of the the boundary of this submanifold proves the desired equation.
The proof of this follows the same line of reasoning as the proof in \cite{FOOO} that bulk deformation :
   
   $$ H^*(X) \to \mathbf{HH}^*(Fuk(X))$$
   
 is a ring homomorphism.\vskip 10 pt
 
\begin{rem} More generally, in the formalism above, if $\mathcal{X}$ is a projective variety and $\mathcal{D}$ is a smooth ample divisor, one could consider chains $S$ in the divisor $\mathcal{D}$ which represent classes of $H_*(\mathcal{D})$ and require that the point of tangency simultaneously lie in $S$. We will explore these deformations of $Fuk(\mathcal{X-D})$ in our forthcoming paper.\end{rem}

To finish the theorem, we have a ``finite determinacy" lemma:
 \begin{lem} The $A(\infty)$ structure on $HF^*_{\mathcal{X}_{(\mathcal{D},d)}}(\mathcal{L},\mathcal{L}) \cong \mathbb{C}[e]/e^{n+1}((t))$ is determined by the fact that $m_j=0$, $2<j<2d$ and $m_{2d}(e^{a_1},e^{a_2},\ldots,e^{a_{2d}})=t$, if $\sum (a_i)=(n+1)d$ \end{lem} 

By a Kunneth theorem, we can get similar results for manifolds of the form $\mathcal{Q}=\prod \mathbb{C}P^{n_j}$. \vskip 10 pt

It seems interesting to make a closer connection between the symplectic geometry in this section and the rational homotopy theory/deformation theory of the previous section. In view of this, it is useful to note the following strong result due to McLean from a recent paper \cite{McLean}.
  \begin{thm} If $T^*\mathcal{Q}$ is symplectomorphic to an affine variety A, then $\mathcal{Q}$ is (rationally) elliptic.\end{thm}

\end{document}